\documentclass[jkps,preprint,fleqn]{revtex4} 
\usepackage{graphicx}
\usepackage{amssymb}
\usepackage{amsmath}
\usepackage{bm}

\pdfoutput=1

\usepackage{float} \usepackage{graphicx}
\usepackage{epstopdf}
\usepackage{graphics}
\usepackage{caption}
\usepackage{subfigure}
\usepackage{epsfig}
\usepackage{color}
\usepackage{multirow,array}

\usepackage{tensor}

\usepackage{adjustbox}  

\usepackage{tabularray}

\usepackage{hyperref}
\hypersetup{
	colorlinks=true,
	linkcolor=blue,
	filecolor=magneta,      
	urlcolor=blue,
}

\begin{document}
\setcounter{page}{0}
\title[]{Review on the minimally extended varying speed of light model}
\author{Seokcheon \surname{Lee}}
\email{skylee@skku.edu}
\affiliation{Department of Physics, Institute of Basic Science, Sungkyunkwan University, Suwon 16419, Korea}

\date[]{Received }

\begin{abstract}
It has been known that dimensional constants such as $\hbar$, $c$, $G$, $e$, and $k$ are merely human constructs whose values and units vary depending on the chosen system of measurement. Therefore, the time variation of dimensional constants lacks operational significance due to their dependence on them. It is well-structured and represents a valid discussion. However, this fact only becomes a meaningful debate within the context of a static or present universe. As well-established theoretically and observationally, the current universe is undergoing accelerated expansion, wherein dimensional quantities, like the wavelength of light, also experience redshift phenomena elongating over cosmic time. In other words, in an expanding universe, dimensional quantities of physical parameters vary with cosmic time. From this perspective, there exists the possibility that dimensional constants, such as the speed of light, could vary with the expansion of the universe. In this review paper, we contemplate under what circumstances the speed of light may change or remain constant over cosmic time, and discuss the potential for distinguishing these cases observationally.
\end{abstract}



\maketitle


\section{Introduction}

The laws of physics should be invariant under changes in units or measurement tools. It is achievable when expressed with dimensionless quantities like the fine structure constant, $\alpha$, as seen in the Standard Model of particle physics. Dimensional constants such as $\hbar$, $c$, $G$, $e$, and $k$ are human constructs whose values vary with the choice of units. In this sense, only dimensionless constants are fundamental. Thus, the potential time variation of dimensionless fundamental constants is a valid subject of inquiry, but that of dimensional constants like $c$ or $G$ is unit-dependent and may lead to disagreement among observers  \cite{Duff:2001ba,Uzan:2002vq,Ellis:2003pw,Duff:2014mva}. However, the above arguments hold only within the context of a static universe or the one at the present epoch \cite{Lee:2020zts,Lee:2023FoP}. 

The contemporary standard cosmological model, known as the $\Lambda$CDM model, is based on the Robertson-Walker (RW) metric, which assumes spatial homogeneity and isotropy on large scales (\textit{i.e.}, cosmological principle (CP)). Evidence for isotropy is found in the uniformity of the cosmic microwave background (CMB) temperature \cite{Hinshaw:2013,Planck:2018nkj}. Although proving homogeneity is more challenging, support comes from the uniform matter distribution on scales of more than 100 million light-years as large-scale structures (LSS) \cite{Guzzo:2018xbe,DES:2020sjz}. The $\Lambda$CDM model incorporates an expanding metric space, evidenced by the redshift of spectral lines in light from distant galaxies. This expansion causes objects not under shared gravitational influence to move apart, but it does not increase their size, such as galaxies. The cosmological redshift, often explained as the result of photon wavelengths stretching due to space expansion, can be understood using equations from general relativity (GR) describing a homogeneous and isotropic universe. This redshift, formulated as a function of the time-varying cosmic scale factor $a(t)$, yields positive values for $z$ in our expanding universe. This phenomenon causes distant galaxies to exhibit redshift as time advances, where $1+z = 1/a$, utilizing the present value of the scale factor as $a_0 = 1$. Therefore, the redshift of a galaxy can be estimated by examining the emission lines emitted by glowing gas within the galaxy. For instance, the H$\alpha$ line, a red Balmer line of neutral hydrogen, has a rest wavelength of $6562$\AA. If the observed wavelength of this line presently measures $8100$\AA, it indicates that the galaxy is positioned at $z = 0.234$ (\textit{i.e.}, $a = 0.81$).  Therefore, in an expanding universe, the value of a dimensional quantity, such as wavelength, \textit{does} vary depending on the time of observation (\textit{i.e.}, cosmic time).  Additionally, it has been observed that the temperature of the CMB decreases with the age of the universe, scaling inversely with the scale factor $T = T_0 a^{-1}$.

The Lorentz transformation (LT) between inertial frames (IFs) stems from special relativity (SR), which hinges on the speed of light, $c$, as its sole parameter with a constant value. SR's universal Lorentz covariance, rooted in Minkowski spacetime, adequately satisfies its principles \cite{Morin07}. In contrast, in GR, an IF refers to a freely falling one. While Lorentz invariant (LI) spacetime intervals can be established between events, defining a global time in GR is hindered by the absence of a universal IF. However, a global time can be defined for the universe satisfying CP, allowing for a foliation of spacetime into non-intersecting spacelike 3D surfaces. It is the universe described by the RW metric \cite{Islam01,Narlikar02,Hobson06,Roos15}. The LI varying speed of light (VSL) model is feasible if $c$ remains locally constant (i.e., at each given epoch) but varies on cosmic time \cite{Lee:2020zts,Lee:2023FoP}. In other words, in an expanding universe, if the speed of light is given as a function of the scale factor, $c[a]$, then although its value varies like wavelengths at different epochs, say $a_1$ and $a_2$, it attains a constant local value at each epoch, ensuring LI and thus maintaining the validity of quantum mechanics and electromagnetism satisfying SR every epoch.  However, testing simultaneous variations in $c$ and Newton’s gravitational constant $G$ is crucial to prevent trivial rescaling of units, given their combination in the Einstein action \cite{Lee:2020zts,Barrow:1998eh}.  Beyond models like meVSL, there exist frameworks in which physical constants vary with cosmic time, one of which is known as Co-varying Physical Couplings (CPC) \cite{Cuzinatto:2022mfe,Cuzinatto:2022vvy,Cuzinatto:2022dta}. The CPC framework presents a modified gravity scenario where the EFEs are assumed but with the quantities ${G, c, \Lambda}$ treated as functions of spacetime. The interplay between the Bianchi identity and the requirement of stress-energy tensor conservation complicates the potential variations of the couplings ${G, c, \Lambda}$, which are compelled to co-vary according to the General Constraint (GC). This model is different from the meVSL by including the dynamics of physical constants by adopting GC. 

The RW metric starts by positing that all galaxies exist on a hypersurface, where the surface of simultaneity of their local Lorentz frame (LF) aligns with this hypersurface. This conceptualization allows for the hypersurface to be visualized as a composite of the smoothly meshed LF of all galaxies, with each galaxy's four-velocity being orthogonal to the hypersurface \cite{Islam01,Narlikar02}. Assigning a parameter $t$ to this sequence of hypersurfaces serves as the proper time of any galaxy, establishing a universal time reference. This cosmic time corresponds to the measurement by a comoving observer, who perceives the universe expanding uniformly around her. Therefore, in the RW metric, the proper time is equivalent to cosmic time \cite{Islam01,Robertson:1929,Walker:1937}.

In the traditional RW metric, the assumption of the constancy of the speed of light is contingent upon a specific hypothesis regarding cosmological time dilation (TD), rather than being directly derived from the foundational principles of the metric \cite{Lee:2023rqv,Lee:2023ucu,SLee:2024}. There have been several projects to measure cosmological TD. Direct observation of the TD measures the decay time of distance supernova (SN) light curves and spectra \cite{Leibundgut:1996qm,SupernovaSearchTeam:1997gem,Foley:2005qu,Blondin:2007ua,Blondin:2008mz}. Another method is measuring TD by searching the stretching of peak-to-peak timescales of gamma-ray bursters (GRBs) \cite{Norris:1993hda,Wijers:1994qf,Band:1994ee,Meszaros:1995gj,Lee:1996zu,Chang:2001fy,Crawford:2009be,Zhang:2013yna,Singh:2021jgr}. There has been a search for the TD effect in the light curves of quasars (QSOs) located at cosmological distances \cite{Hawkins:2001be,Dai:2012wp}.  So far, it seems fair to say that no convincing detection has been made for cosmic time dilation with the conflict between different measurements.  Without explicit laws governing TD, the speed of light in the RW metric can vary with cosmological time, similar to other physical properties such as mass density, temperature, and fundamental constants like the Planck constant \cite{Lee:2022heb}. This variation presents a plausible scenario known as the VSL with cosmic time. When delineating the background of the Friedmann-Lema\^itre-Robertson-Walker (FLRW) universe, a hypersurface of constant time can be delineated based on physical quantities such as temperature or density, owing to the Universe's homogeneity, which ensures uniform temperature and density at each cosmic time. Nevertheless, it is crucial to recognize that temperature and mass density can undergo redshift due to the Universe's expansion. 

In Section~\ref{sec:RWmetric}., we review the conventional derivation of the RW metric using CP principles and Weyl’s postulate. Section~\ref{sec:FLRW} explores the possibility of the VSL model in the RW metric by considering the various implications of TD relationships. Subsequently, Section~\ref{sec:ConVSL} comprehensively investigates the repercussions of VSL on a range of physical quantities and fundamental constants. In Section~\ref{sec:FriEq}, we compare the Friedmann equations of various models.  We address observational methods that can distinguish the minimally extended VSL (meVSL) model from the standard model cosmology (SMC) in Section~\ref{sec:Obs}. In Section~\ref{sec:Conc}, we will summarize the main points and draw conclusions based on the insights gained throughout the document, emphasizing the potential ramifications of the cosmological time-varying speed of light across various dimensions of its definitions and implications.

\section{Summary of the Robertson-Walker Metric}
\label{sec:RWmetric}

The contemporary standard model of cosmology, the $\Lambda$CDM model, relies on the RW metric, which assumes spatial homogeneity and isotropy in the Universe on its largest scales. Through a synthesis of observational data from LSS and CMB, there is a consensus that the Universe demonstrates nearly perfect homogeneity and isotropy on a large scale. In this section, we offer a comprehensive examination of the derivation of the RW metric, employing the CP and Weyl’s postulate.

\subsection{Isotropic and Homogeneous space}
\label{subsec:IHspace}

First, it's crucial to highlight that the isotropy and homogeneity of space are defined at each moment in time. We present a review of a method utilizing Killing vectors (KVs) to translate the CP into a geometric condition explicitly satisfied by the spacetime metric, particularly its spatial component \cite{Ryder09}. Initially, we note that the Lie derivatives of the KVs, responsible for generating isotropy symmetry in a $3$-dimensional space (\textit{i.e.}, the metric remains invariant under spatial rotations), must vanish. These three KVs are in spherical polar coordinates $x^{\mu} = (ct',r,\theta,\phi)$ as 
\begin{align}
&\bar{\xi}_{i} = \xi_{i (\text{sp})}^{\mu} \bar{e}^{(\text{sp})}_{(\mu)} \, , \text{where} \,\,\, \bar{e}^{(\text{sp})}_{(\mu)}  = \left( \frac{\partial}{c \partial t}, \frac{\partial}{\partial r}, \frac{\partial}{\partial \theta}, \frac{\partial}{\partial \phi} \right) \,, \xi_{1(\text{sp})}^{\mu} = \left( 0, 0, -\sin \phi, -\cot \theta \cos \phi \right)  \nonumber \,, \\
&\xi_{2(\text{sp})}^{\mu} = \left( 0\,, 0\,, \cos \phi \,, -\cot \theta \sin \phi \right) \, ,  \, \xi_{3(\text{sp})}^{\mu} = \left( 0\,,0\,, 0 \,, 1 \right) \,.  \label{IsoKVs}  
\end{align} 

Therefore, by applying these KVs to the condition of rotational isometry (${\cal L}_{\bar{\xi}} g_{\mu\nu} = 0$), we derive the most comprehensive metric component for an isotropic space at a specific time, represented as $t' = t_{l}'$
\begin{align}
g_{\mu\nu}^{(\text{iso})}(t_l') = \begin{pmatrix} g_{00}(t_{l}'\,,r) & g_{01}(t_{l}'\,,r) & 0 & 0 \\ 
g_{10}(t_{l}'\,,r) & g_{11}(t_{l}'\,,r) & 0 & 0 \\ 0 & 0 & g_{22}(t_{l}'\,,r) & 0 \\ 0& 0 & 0 & g_{33}(t_{l}'\,,r) \end{pmatrix}  \label{gmunuiso} \,.
\end{align}
On cosmological scales,  space is also homogeneous, so the above metric must satisfy the translation isometry for three translation KVs as given by  
\begin{align}
&\bar{\eta}_{i} = \eta_{i(\text{sp})}^{\mu} \bar{e}^{(\text{sp})}_{(\mu)}\,, \eta_{1 (\text{sp})}^{\mu} = \left( 0, \sin \theta \cos \phi,  \frac{1}{r} \cos \theta \cos \phi,  -\frac{1}{r} \frac{\sin \phi}{\sin \theta}  \right)   \nonumber \,, \\
&\eta_{2 (\text{sp})}^{\mu} = \left( 0, \sin \theta \sin \phi,   \frac{1}{r} \cos \theta \sin \phi,  \frac{1}{r} \frac{\cos \phi}{\sin \theta} \right) \,,  \eta_{3 (\text{sp})}^{\mu} = \left( 0, \cos \theta,  -\frac{1}{r} \sin \theta,  0 \right) \,.  \label{homKVs} 
\end{align} 
The most general metric component for an isotropic and homogenous space at $t_l'$is obtained by employing these KVs to the translation isometry for the isotropic metric provided in \eqref{gmunuiso} ({\it i.e.},  ${\cal L}_{\bar{\eta}} g_{\mu\nu}^{(\text{iso})} = 0$)
\begin{align}
g_{\mu\nu}^{(\text{CP})}(t_l') = \text{diag} \left( g_{00} (t'_l) \,,  A(t'_l) \,,  A(t'_l) r^2 \,,  A(t'_l) r^2 \sin^2 \theta \right) \,. \label{gmunuCP}
\end{align}
Consequently, the general form of the 4-dimensional line element for the homogeneous and isotropic space at a particular time $t_l'$ can be expressed as 
\begin{align}
ds^2(t_l') = g_{\mu\nu}(t'_l) dx^{\mu} dx^{\nu} = -c_l^2 g_{00}(t'_l) dt_l^{'2} + A(t'_l) \left[ dr^2 + r^2 d \theta^2 + r^2 \sin^2 \theta d \phi^2  \right]  \label{dsl21}  \,.
\end{align}
By introducing the transformation $dt_{l} = \sqrt{g_{00}(t'_l)} dt'_l$ (\textit{i.e.}, normalizing the lapse function as $1$),  the homogeneous and isotropic metric in \eqref{gmunuCP} can be rewritten as
\begin{align}
g_{\mu\nu}(t_l) = \text{diag} \left( -1,  a^2(t_l),  a^2(t_l) r^2,  a^2(t_l) r^2 \sin^2 \theta \right) \label{gmunuCPN} \,,
\end{align}
where $a^2(t_l) = A(t'_l)$.  Consequently,  the $4$-dimensional line element for the homogeneous and isotropic space at a particular time $t_l$ is given by
\begin{align}
ds_l^2 &\equiv ds^2(t_l) = -c_l^2 dt^2 + a^2(t_l) \left[ dr^2 + r^2 d \theta^2 + r^2 \sin^2 \theta d \phi^2  \right] \nonumber \\ 
&\equiv -c_l^2 dt^2 + a^2(t_l) \left[ dr^2 + r^2 d \Omega^2  \right] \label{dsl2}  \,.
\end{align}
It is crucial to emphasize that both the scale factor $a_l$ and the speed of light $c_l$ in Eq. ~\eqref{dsl2} must remain constant to maintain homogeneity at a specific time $t_l$.

\subsection{Fundamental Observers}
\label{subsec:FundObs}
In the preceding subsection \ref{subsec:IHspace}, we derived the most general metric for an isotropic and homogeneous space at a specific time, denoted as $t_l$. To effectively utilize this metric in cosmology, it becomes necessary to introduce a global time parameter (\textit{i.e.}, referred to as generalized $t_l$ as $t$). In SR, one can define a globally valid time within a selected IF. In GR, the absence of a global IF renders the concept of an instantaneous moment ambiguous. Instead,  three-dimensional spacelike hypersurfaces replace this notion. By introducing non-intersecting spacelike hypersurfaces labeled by $t_l$s, a global time parameter can be defined, representing a universal time \cite{Islam01,Hobson06}. 

To specify the preferred slicing, fundamental observers, assumed to have no motion relative to the overall cosmological fluid, are introduced. If Weyl's postulate is adopted,  the timelike worldlines of these observers form a bundle in spacetime, diverging from or converging to a point in the past or future.  Thus, $t_l =$ constant hypersurfaces are constructed where the four-velocity of any fundamental observer is orthogonal to it.  Consequently, each hypersurface can be seen as the amalgamation of all the local LFs of fundamental observers. \cite{Islam01}. 

In relativistic cosmology, Weyl’s postulate dictates that the worldlines of fluid particles ({\it i.e.},  galaxies) should be hypersurface orthogonal, meaning they should be everywhere orthogonal to a family of spatial hyperslices. In essence, fundamental observers move with the cosmic fluid, remaining at rest in the comoving frame.

\subsection{Synchronous coordinates}
\label{subsec:Syncoord}
The parameter $t_l$ assigned to hypersurfaces is termed the {\it synchronous time coordinate} when considered as the proper time along the worldline of any fundamental observer \cite{Islam01,Narlikar02,Hobson06,Ryder09}.  Additionally,  the so-called comoving coordinates can be adopted, where each fundamental observer possesses fixed spatial coordinates $x^{i} = (x^1 \,, x^2 \,, x^3 )$.  Consequently, the worldline of a fundamental observer is expressed as $x^{\mu}(\tau)$ 
\begin{align}
x^{\mu}(\tau) = \left( x^0 \,, x^i \right) = \left( c_l \tau_l \,, x^1 = \text{constant} \,,  x^2 = \text{constant} \,,  x^3 = \text{constant} \right) \,, \label{xmutau} 
\end{align}
where $\tau_l$ represents the proper time along the fundamental observer.  Since $dx^i = 0$ along the worldline,  this yields $ds_l = c_l d \tau_l = c_l dt_l$. Hence,  the proper time along the worldline equals the coordinate time $t_l$,  commonly referred to as cosmic time \cite{Islam01,Hobson06}. The four-velocity of a fundamental observer in comoving coordinates, denoted as $u^{\mu} \equiv dx^{\mu}/d\tau_l= (c_l  \,, 0 \,, 0 \,, 0 )$, is orthogonal to any vector $B^{\mu} = ( 0,  dx^1 \,, dx^2 \,, dx^3 )$ lying in the hypersurface $t_l = $ constant,  meaning $g_{\mu\nu} u^{\mu} B^{\nu} = 0$.     

Let us consider two nearby fundamental observers located at comoving coordinates $(x^1,  x^2,  x^3 )$ and $(x^1 + d x^{1},  x^2 + d x^{2},  x^3 + d x^{3} )$ at a specific time $t = t_l$.  In the flat-space case, their proper (physical) distances from the origin along the Cartesian coordinate axes are given at this time by
\begin{align}
&(r^1,  r^2,  r^3 ) = a(t_l) (x^1,  x^2,  x^3 )  ,  \nonumber \\ 
&(r^1 + \Delta r^{1},  r^2+ \Delta r^{2},  r^3+ \Delta r^{3} ) = a(t_l) (x^1 + d x^{1},  x^2 + d x^{2},  x^3 + d x^{3} ) \,. \label{r123}
\end{align} 
Considering the triangles formed by these observers at $t_l$ and at some later time, due to homogeneity and isotropy of space, both triangles must be similar. the triangle formed by these same observers at some later \cite{Islam01}.  Moreover, the magnification factor must be constant regardless of the triangle's position in the three-dimensional space.  Therefore, the spatial separation on the same hypersurface $t = t_l$ between two nearby fundamental observers is expressed as
\begin{align}
d \sigma_{l}^2 \equiv g_{ij}(t_l) \Delta r^i \Delta r^j = a^2(t_l) \gamma_{ij} dx^i dx^j \equiv a^2(t_l) dl^2 \label{dsigmal2} \,,
\end{align}
where the $\gamma_{ij}$ depends solely on $(x^1,  x^2,  x^3 )$ and becomes $\textrm{diag}(1,1,1)$ in the flat space \cite{Islam01,Hobson06}. 

\subsection{Curved spatial hypersurface}
\label{subsec:CSH}
The spatial section in the line element in Eq.\eqref{dsl2} is flat, as it is derived using translational isometry in subsection \ref{subsec:IHspace}. A curved spatial hypersurface can be embedded into a flat $4$-dimensional Euclidean space $\mathbb{E}^4$. Closed and open spaces correspond to spherical and hyperbolic spaces, respectively, with a curvature radius of $b$
\begin{align}
&\left( x^1 \right)^2 + \left( x^2 \right)^2 + \left( x^3 \right)^2 + \left( x^4 \right)^2 \equiv r'^2 + \left( x^4 \right)^2 = b^2,\text{where} \label{S3}\\
&x^1 = b \sin \chi \sin \theta \cos \phi \, , \, x^2 = b \sin \chi \sin \theta \sin \phi \, , \,  x^3 = b \sin \chi \cos \theta \, ,  \, x^4 = b \cos \chi  \,,  \nonumber  \\
&- \left( x^1 \right)^2 - \left( x^2 \right)^2 - \left( x^3 \right)^2 + \left( x^4 \right)^2 \equiv -r'^2 + \left( x^4 \right)^2= b^2, \text{where}  \label{H3} \\
&x^1 = b \sinh \chi \sin \theta \cos \phi \, , \, x^2 = b \sinh \chi \sin \theta \sin \phi \, , \,  x^3 = b \sinh \chi \cos \theta \, ,  \, x^4 = b \cosh \chi \nonumber \,.
\end{align}
To consider a curved space,  three translation KVs given in Eq.~\eqref{homKVs} should be replaced with KVs for rotations along the $x^4$ axis \cite{Ryder09} 
\begin{align}
\bar{\eta}_1 = x^4 \frac{\partial} {\partial x^1} - x^1 \frac{\partial}{\partial x^4} \,,\, \bar{\eta}_2 = x^4 \frac{\partial} {\partial x^2} - x^2 \frac{\partial}{\partial x^4}  \,,\, \bar{\eta}_3 = x^4 \frac{\partial} {\partial x^3} - x^3 \frac{\partial}{\partial x^4}  \label{KVsRot2} \,.
\end{align} 
Then,  the line element for the homogeneous and isotropic spaces at $t = t_l$ for maximally symmetric closed $3$-space ($\mathbb{S}^3$) can be expressed 
\begin{align}
ds_l^2 &= - c_l^2 dt_l^2 + a_l^2 \left( \frac{dr'^2}{1- K r'^2} + r'^2 \left( d \theta^2 + \sin^2 \theta d \phi^2 \right)  \right)\,,\text{where} \quad K = \frac{1}{b^2} \,. \label{dsl2close} 
\end{align}
Also, the line element for the open $3$-space (hyperbolic) is expressed as 
\begin{align}
ds_l^2 &= - c_l^2 dt_l^2 + a_l^2 \left( \frac{dr'^2}{1- K r'^2} + r'^2 \left( d \theta^2 + \sin^2 \theta d \phi^2 \right)  \right)\,,\text{where} \quad K = - \frac{1}{b^2} \,. \label{dsl2open} 
\end{align}


Therefore, the spacetime interval for a curved spatial hypersurface adhering to the CP at a specific time $t_l$ can be expressed by 
\begin{align}
ds_{l}^2 = - c_l^2 dt^2 + a_l^2 \left[ \frac{dr^2}{1-Kr^2} + r^2  \left( d \theta^2 + \sin^2 \theta d \phi^2 \right)  \right] \quad \textrm{at} \,\,  t= t_l \label{dststar} \,. 
\end{align}

\subsection{Robertson Walker metric}
\label{subsec:RWm}

\begin{figure}
	\begin{center}
	\includegraphics[width=0.9\textwidth]{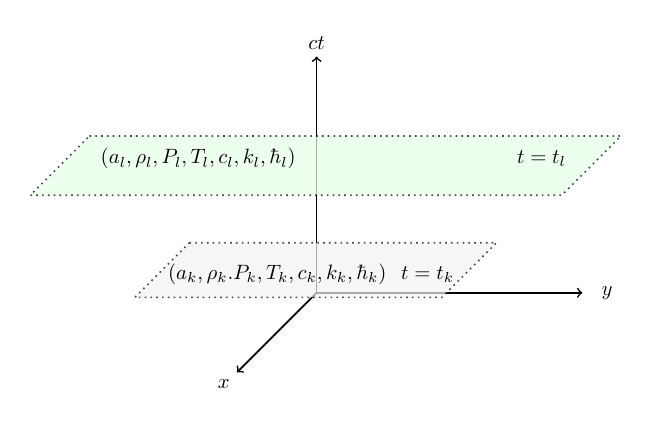} 
	\caption{At $t = t_k$, the values of physical quantities and constants, such as $a_k$, $\rho_k$, $P_k$,  $T_k$, $c_k$,  $k_{k}$, and $\hbar_k$, are fixed and independent of spatial position on the $t=t_k$ hypersurface. As the universe expands, these quantities and constants transition to $a_l$, $\rho_l$, $P_l$,  $T_l$, $c_l$,  $k_l$, and $\hbar_l$. The CP and Weyl’s postulate do not restrict $c_k$ to be equal to $c_l$; its value is determined by the cosmological TD relation. }
	\label{Fig1}
	\end{center}
\end{figure}

By adopting Weyl's postulate to extend the metric described in Eq.~\eqref{dststar} to cosmic time $t$, we can represent the line element as
\begin{align}
ds^2 = - c(t)^2 dt^2 + a(t)^2 \left[ \frac{dr^2}{1-Kr^2} + r^2  \left( d \theta^2 + \sin^2 \theta d \phi^2 \right)  \right] \equiv - c(t)^2 dt^2 + a(t)^2 dl_{3\textrm{D}}^2 \label{dstgen} \,. 
\end{align} 
In this equation \eqref{dstgen},  the speed of light is expressed as a function of time, deviating from the conventional RW metric. Initially, this equation may seem incorrect or counterintuitive. However, as illustrated in Figure~\ref{Fig1}, the original RW metric implies that on hypersurfaces defined by $t_l$ or $t_k =$ constants, various quantities such as the scale factor $a_l = a(t_l)$, mass density $\rho_l = \rho(t_l)$,  pressure $P_l = P(t_l)$, temperature $T_l = T(t_l)$, speed of light $c_l = c(t_l)$,  Boltzmann constant $k_{l} = k(t_l)$, and Planck constant $\hbar_l = \hbar(t_l)$ remain constant regardless of the $3$-D spatial position.  However, according to Weyl’s postulate, these quantities or constants can be expressed as functions of cosmic time $t$, accounting for cosmological redshift, as depicted in Figure \ref{Fig1}. Traditionally, it has been assumed that physical constants, including the speed of light, remain constant over cosmic time. This additional assumption, namely that the speed of light remains constant ($c_l = c_k$) regardless of cosmic time, is not directly tied to the two conditions necessary to derive the RW metric: the CP and Weyl’s postulate. The constancy of the speed of light relies on cosmological TD, and it is crucial to recognize that GR does not specify any particular physical laws governing this constancy, as we will elucidate shortly. As the Universe progresses from $t_k$ to $t_l$, physical quantities such as $a(t)$, $\rho(t)$, $P(t)$, and $T(t)$ undergo changes over cosmic time $t$. The precise functional expressions for these quantities are obtained through the solution of Einstein's Field Equations (EFEs) and Bianchi's identity (BI), considering the equation of state of fluids \cite{Lee:2020zts,Lee:2023FoP}.

\subsection{Rescale time}
\label{subsec:ResT}

Some argue that debating whether the speed of light varies is meaningless when transitioning from equation~\eqref{dststar} to equation~\eqref{dstgen} by substituting $ct$ with $x^{0}$. However, as demonstrated in equation~\eqref{dl3D}, when defining the $3$D comoving distance that light traverses along its path, this equation becomes 
\begin{align}
d l_{3\textrm{D}} &= \frac{x^{0}(t_i)}{a(t_i)} \quad \Rightarrow \quad \frac{x^{0}(t_1)}{a_1} = \frac{x^{0}(t_2)}{a_2} \quad \Rightarrow \quad \frac{\lambda_1}{a_1} = \frac{\lambda_2}{a_2} \,, \label{dl3Dapp}
\end{align}
where $x^0$ represents distance related to the wavelength of light. Thus, this equation illustrates the cosmological redshift of the wavelength, mirroring Eq. ~\eqref{dl3D2}. Dividing it by the clock rate $\nu$ and the speed of light $c$ reveals both possibilities: the speed of light either remains constant or varies with cosmic time, contingent upon the cosmological TD outlined in Eq. ~\eqref{dl3D}.

\section{The Possibility of Varying Speed of Light Theory in the Robertson-Walker Metric}
\label{sec:FLRW}

The derivation of redshift involves employing the geodesic equation for a light wave, where $ds^2 = 0$ as Eq. ~\eqref{dstgen}. The consistency of $dl_{3\textrm{D}}$ over time is ensured by the exclusive use of comoving coordinates. Expanding upon this groundwork, we reach the expression for radial light signals as
\begin{align}
d l_{3\textrm{D}} &= \frac{c(t_i) dt_i}{a(t_i)} \quad : \quad \frac{c_1 dt_1}{a_1} = \frac{c_2 dt_2}{a_2} \Rightarrow \begin{cases} c_1 = c_2 = c & \textrm{if} \quad \frac{dt_1}{a_1} = \frac{dt_2}{a_2} \qquad \textrm{SMC} \\ 
 c_1 = \frac{f(a_2)}{f(a_1)} \frac{a_1}{a_2} c_2 & \textrm{if} \quad \frac{dt_1}{f(a_1)} = \frac{dt_2}{f(a_2)} \quad \textrm{VSL} \\ c_1 = \left( \frac{a_1}{a_2}\right)^{\frac{b}{4}} c_2 & \textrm{if} \quad \frac{dt_1}{a_1^{1-\frac{b}{4}}} = \frac{dt_2}{a_2^{1-\frac{b}{4}}} \quad \textrm{meVSL}  \end{cases} \,, \label{dl3D}
\end{align}
where  $d t_i = \nu(t_i)$ represents the time interval between successive crests of light at $t_i$ (\textit{i.e.}, the inverse of the frequency $\nu_i$ at $t_i$), and $f(a_i)$ denotes an arbitrary function of $a(t_i)$ \cite{Weinberg:2008}.

In the SMC, an extra assumption is made, asserting the constancy of the speed of light as $c$. It stems from SMC's reliance on GR, where $c$ is regarded as a constant. Consequently, the cosmological TD between two hypersurfaces at $t = t_1$ and $t = t_2$ is directly related to the inverse of the scale factors $a(t)$ at those specific times. However, it lacks derivation from any physical laws. In contrast to this assumption, if the speed of light varies with time, as hypothesized in this paper, this relationship may no longer hold.

Conversely, in an expanding universe, the progression from one hypersurface to another results in an increase in the scale factor, naturally leading to the cosmological redshift of various physical quantities, including mass density and temperature. However, it is impossible to conclude about cosmological TD based solely on the CP and Weyl’s postulate in the RW metric. Instead, establishing such relationships relies on experimental observations. Efforts to measure cosmological TD have included direct observations of SN light curves and spectra to evaluate decay times of distance \cite{Lee:2023ucu,Leibundgut:1996qm,SupernovaSearchTeam:1997gem,Foley:2005qu,Blondin:2008mz}. Another avenue to explore cosmological TD involves analyzing the elongation of peak-to-peak timescales observed in GRBs \cite{Norris:1993hda,Wijers:1994qf,Band:1994ee,Meszaros:1995gj,Lee:1996zu,Chang:2001fy,Crawford:2009be,Zhang:2013yna,Singh:2021jgr}. Additionally, researchers have investigated TD effects within the light curves of cosmologically distant QSOs \cite{Hawkins:2001be,Dai:2012wp,Lewis:2023jab}.  However, current observational evidence does not definitively confirm an exact correspondence between cosmological TD and predictions made by the SMC. Moreover, the RW model lacks a mechanism to determine cosmological TD conclusively. Thus, it remains valuable to explore the possibility of VSL in these observations, provided that the findings are consistent with those predicted by the SMC.

Given the theoretical absence of cosmological TD, considering this relationship as a general function $f(a)$ of the scale factor, the speed of light can be expressed as 
\begin{align}
c(t_1) = \frac{f(a_2)}{f(a_1)} \frac{a(t_1)}{a(t_2)} c(t_2) \label{cVSL} \,.
\end{align}
This underscores that while we cannot assert the generality of the VSL model within the framework of GR, it appears to be a natural consequence in an expanding Universe as described by the RW metric. The meVSL model is a specific instance of VSL, characterized by $f(a) = a^{1-b/4}$ \cite{Lee:2020zts,Lee:2023FoP}.

\section{The consequences of the varying speed of light}
\label{sec:ConVSL}

In the previous section~\ref{sec:FLRW}, we illustrated how the speed of light may change over cosmic time within an expanding Universe according to the RW metric. However, for this concept to form a coherent model, the variable nature of the speed of light must be integrated into EFEs and solved for solutions. Our previous works have addressed such scenarios, particularly in the context of a model known as meVSL \cite{Lee:2020zts,Lee:2023FoP}.  In this section, we delve into extending the cosmological evolution of physical quantities and constants to encompass a broader range of VSL models.

\subsection{Stress Energy Tensor}
\label{subsec:EMT}

In cosmology, matter is treated as a perfect fluid, characterized by its total mass density $\rho$ and isotropic pressure $P$. $\rho$ contains both the rest-mass density measured in the fluid's rest frame and the mass content of the internal elastic energy density. In the preceding subsection \ref{subsec:FundObs}, we introduced the notion of a fundamental observer at rest relative to this fluid. Within the framework of GR, the stress-energy tensor describes this perfect fluid
\begin{align}
T^{\mu\nu} = \left( \rho + \frac{P}{c^2} \right) u^{\mu} u^{\nu} + P g^{\mu\nu} \label{Tmunu} \,,
\end{align}
where $u^{\mu}$ represents its four-velocity. When the fluid is in motion, a set of fundamental observers is deemed comoving with it, characterized by a four-velocity denoted as $u^{\mu} = (c, 0,0, 0)$, as discussed in section \ref{subsec:Syncoord}. Once we have established the metric and the stress-energy tensor, the subsequent step entails solving EFEs to elucidate the dynamics of the scale factor in the metric. These equations govern the dynamics of expansion, encompassing the speed and acceleration of the Universe's expansion as observed between two fundamental observers.

\subsection{Cosmological redshift in the RW metric}
\label{subsec:redshift}

We obtain the cosmological redshift through the geodesic equation to a light wave ($ds^2 = 0$). When analyzing electromagnetic waves traversing through a vacuum, it is essential to highlight the maintenance of a linear dispersion relation, expressed as $\lambda_i \nu_i = c_i$.  It is noteworthy that the SMC, VSL, and meVSL models all yield identical predictions for a given wavelength's cosmological redshift, as depicted in Eq. ~\eqref{dl3D}
\begin{align}
\frac{c(t_i) d t_i}{a(t_i)} \equiv \frac{c_1}{a_1 \nu_1} = \frac{c_2}{a_2 \nu_2} \quad 
\Rightarrow \quad \lambda_1 = \frac{a_1}{a_2} \lambda_2 \,. \label{dl3D2}
\end{align} 
Determining the redshift of a distant source entails scrutinizing its light spectrum, particularly absorption or emission lines, and variations in light intensity. Cosmological redshift, attributed to the Universe's expansion, is defined by the relative discrepancy between the wavelengths observed and emitted by an object \cite{Weinberg:2008}. By observing redshifted wavelengths, we can gain insights into the characteristics of photons beyond their speed and frequency. Consequently, VSL models find a natural framework within the RW metric, especially during periods of universal expansion.

In this manuscript, we limit our consideration of the cosmological redshift to the Planck relation which states that the energy of a photon ($E$) is given by
\begin{align}
 	E(a) = h \nu = h \frac{c}{\lambda}  =
		\left.
	  \begin{cases}
		 h_0 a^{-b/4} \frac{c_0 a^{b/4}}{\lambda_0 a} = h_0 \frac{c_0}{\lambda_0} a^{-1} & \text{meVSL} \\
		 h_0  \frac{c_0}{\lambda_0 a} = h_0 \frac{c_0}{\lambda_0} a^{-1} & \text{SMC} \\
  \end{cases}
  \right\}
= E_0 a^{-1} \label{Ea} \,,
\end{align}
where $E_0 \equiv E(a=a_0=1)$. We show this in Table I. However, if one worries about the atomic level of energy change from the Rydberg energy level, then one should consider this effect, which can be found in \cite{Lee:2023xfg}.

The energy scale $E_{R}$ in the meVSL model exhibits a dependence on the scale factor $a$, expressed as 
\begin{align}
    E_{R} = \frac{m_{e0} e_{0}^{4}}{2 \left( 4 \pi \epsilon_0 \right)^2 \hbar_0^2} a^{-\frac{b}{2}} \equiv E_{R0} a^{-\frac{b}{2}} \label{ERz} \,.
\end{align}
In the non-relativistic regime, all energy scales of atomic spectra are characterized by the Rydberg unit, $E_R$, with any cosmological evolution being absorbed into the determination of the redshift parameter $z$, given by
\begin{align}
    \lambda_{i}^{(\textrm{non-rel})} \propto \frac{hc}{E_R} = \frac{h_0c_0}{E_{R0}} a^{\frac{b}{2}} \equiv \lambda_{i 0}^{(\textrm{non-rel})} (1+z) \label{lambdainonrel} \,,
\end{align}
where $\lambda_{i 0}^{(\textrm{non-rel})}$ denotes the present (laboratory) value of the wavelength. This would introduce an additional factor in the measurement of $z$ in the meVSL model.

\subsection{Hilbert-Einstein action}
\label{subsec:HEaction}

The EFEs stem from the principle of least action via the Einstein-Hilbert (EH) action
\begin{align}
S &\equiv \int \Biggl[ \frac{1}{2 \kappa} \left( R - 2 \Lambda \right) + \mathcal L_{m} \Biggr] \sqrt{-g} dt d^3x \label{SHmpApp} \,,
\end{align}
where $\kappa$ represents the Einstein gravitational constant. Introducing a variation exclusively in the speed of light as a function of cosmic time presents a challenge in deriving EFEs due to the impact of the Palatini identity term on the varying speed of light. Hence, allowing for a variation in the gravitational constant becomes essential to ensure that the combination of these constants ($\kappa = 8 \pi G/c^4$) within the EH action remains unaffected by cosmic time
\begin{align}
\kappa = \textrm{const} 
\quad \Rightarrow \quad \begin{cases} G_1 = G_2 = G & \textrm{SMC} \\ 
G_1 = \frac{f(a_2)^4}{f(a_1)^4} \frac{a_1^4}{a_2^4} G_2 & \textrm{VSL} \\ G_1 = \left( \frac{a_1}{a_2}\right)^{b} G_2 & \textrm{meVSL}  \end{cases} \,. \label{G}
\end{align} 
EFEs, which incorporate the cosmological constant, can be expressed as 
\begin{align}
&R_{\mu\nu} - \frac{1}{2} g_{\mu\nu} R + \Lambda g_{\mu\nu} \equiv G_{\mu\nu} + \Lambda g_{\mu\nu}  = \frac{8 \pi G}{c^4} T_{\mu\nu} \label{tEFEmpApp} \,, 
\end{align}
where $G_{\mu\nu}$ represents the Einstein tensor. The structure of EFEs closely resembles that of the SMC. 

The Brans–Dicke (BD) theory of gravitation, sometimes referred to as the Jordan–Brans–Dicke (JBD) theory, can be an alternative to Einstein's GR. This theory belongs to the category of scalar-tensor theories, which are gravitational theories wherein the gravitational interaction is governed not only by the tensor field of general relativity but also by a scalar field. In this framework, the gravitational constant $G$ is not a constant; instead, the reciprocal of $G$ is substituted by a scalar field $\phi$, which may vary in both space and time.

If we apply scalar-tensor theories of gravity, we could probably remove the constraints on equation \eqref{G} and obtain dynamical equations for $G$ and $c$. However, in this manuscript, we are addressing all discussions within the framework of the minimal extension theory of general relativity, which we term the meVSL model.

\subsection{Adiabatic expansion and cosmological evolution of the Planck constant}
\label{subsec:AdiabaticExp}

As outlined in Section \ref{sec:FLRW}, the RW metric accommodates cosmological VSL while upholding the CP. Preserving adiabatic conditions is crucial to maintain homogeneity and isotropy. If the speed of light varies over time while leaving other physical constants unchanged, it could potentially disrupt the isotropy of the Universe.  An uneven energy flux might undermine isotropy if there is a preferential direction for energy flow. Adiabaticity also has the potential to promote homogeneity if the outward (or inward) energy flow remains isotropic. For a VSL model reliant on an expanding universe to be considered viable, it must adhere to the requirement of adiabatic expansion. This condition also leads to the cosmological evolution of the Planck constant within VSL models \cite{Lee:2022heb}. 

The first law of thermodynamics embodies energy conservation, illustrating that a fluid element within a momentarily comoving reference frame can exchange energy with its surroundings through heat conduction (absorption of heat) and work (exertion of work). Given the significant contribution of photons to entropy, our focus is primarily on photons. Consequently, the first law of thermodynamics can be expressed as
\begin{align}
&dQ = dE + P dV \equiv d (\varepsilon_{\gamma} V) + P_{\gamma} dV \quad ,\, \textrm{where} \nonumber \\ 
&\varepsilon_{\gamma} = \frac{\pi^2}{15} \frac{\left( k_{\textrm{B}} T_{\gamma} \right)^4}{\left( \hbar c \right)^3} \equiv \sigma_{\gamma} T^4 \quad , \quad  p_{\gamma} = \frac{1}{3} \varepsilon_{\gamma} 
\label{dQ} \,,
\end{align}
where $\sigma_{\gamma}$ represents the black-body constant. Processes in which $dQ = 0$ are termed adiabatic processes, with the adiabatic expansion of the Universe leaving its entropy unaffected. The following equation describes the change in heat concerning the variation of the speed of light
\begin{align}
dQ = 4 \sigma_{\gamma} T_{\gamma}^4 V_0 a^3 \left[ d \ln T_{\gamma} + d \ln a + \frac{1}{4} d \ln \sigma_{\gamma} \right] = 0 \label{VSLS} \,.
\end{align}
To maintain the observed behavior of the CMB temperature's time evolution, $T_{\gamma} = T_{\gamma 0}a^{-1}$, the term $d \ln \sigma_{\gamma}$ must vanish. This observation implies the absence of time evolution for the Boltzmann constant while inducing the reduced Planck constant as
\begin{align}
\sigma_{\gamma} = \textrm{const} 
\quad \Rightarrow \quad \begin{cases} \hbar_1 = \hbar_2 = \hbar & \textrm{SMC} \\ 
 \hbar_1 = \frac{f(a_1)}{f(a_2)} \frac{a_2}{a_1} \hbar_2 & \textrm{VSL} \\ \hbar_1 = \left( \frac{a_2}{a_1}\right)^{\frac{b}{4}} \hbar_2 & \textrm{meVSL}  \end{cases} \,. \label{Planck}
\end{align} 

In reference~\cite{Lee:2022heb} one can put limits on the deviation of the time evolution of $T$ as $T_0 a^{-1+\beta}$ and various datasets from different missions have been utilized to constrain the value of $\beta$, as summarized in Table~\ref{Tab:1}. The South Pole Telescope (SPT) employs measurements of the Sunyaev-Zeldovich effect (SZe) spectrum, obtained at frequencies of $95$ and $150$ GHz, to assess deviations from the expected adiabatic evolution of the cosmic microwave background (CMB) temperature \cite{SPT:2013gam}. This method is applied to a dataset comprising $158$ SPT-selected clusters spanning a redshift range of $0.05 < z < 1.35$. Additionally, Planck temperature maps, covering frequencies from 70 to 353 GHz, are utilized to obtain SZe spectra for a subset of 104 clusters from the Planck SZ cluster catalog. Employing a Monte-Carlo Markov Chain approach and examining SZ intensity changes across different frequencies, individual measurements of CMB temperature are derived for each cluster in the sample \cite{Luzzi:2015via}. Furthermore, utilizing data from $370$ clusters obtained from the largest SZ-selected cluster sample to date, collected by the Atacama Cosmology Telescope (ACT), new constraints on the deviation of CMB temperature evolution from the SMC are derived \cite{Li:2021hxe}. Notably, all these findings are consistent with $ \beta = 0$, indicating adiabatic expansion.

 \begin{table*}[h!]
 	\centering
\begin{tabular}{ |c|c|c|c| } 
 \hline
 & SPT &  Planck DR$1$ & ACT \\ 
\hline
$\beta$ & $0.017_{+0.030}^{-0.028}$ &  $0.012 \pm 0.016$ 
& $0.017_{-0.032}^{+0.029}$ \\ 
\hline
ref & \cite{SPT:2013gam} & \cite{Luzzi:2015via} & \cite{Li:2021hxe} \\ 
\hline
\end{tabular}
\caption{These are the values of $\beta$ obtained from various missions.}
\label{Tab:1}
 \end{table*}

\subsection{Einstein tensors}
\label{subsec:ETs}

We initiate the derivation of Einstein’s tensors for VSL models within the framework of the RW metric, as depicted in Eq. ~\eqref{dstgen}. The Christoffel symbols associated with the RW metric are defined by 
\begin{align}
&\Gamma^{0}_{ij} = \frac{a\dot{a}}{c} \gamma_{ij} \quad , \quad \Gamma^{i}_{0j} = \frac{1}{c}  \frac{\dot{a}}{a} \delta^i_j \quad , \quad \Gamma^{i}_{jk} = ^{s}\Gamma^{i}_{jk}  \label{GammacompApp} \,,
\end{align}
where $^{s}\Gamma^{i}_{jk}$s represent the Christoffel symbols related to the spatial metric $\gamma_{ij}$. In Eq.~\eqref{GammacompApp}, the Christoffel symbols for VSL models take forms akin to those of the SMC. However, in VSL models, the value of $c$ varies with the scale factor. While the Christoffel symbols of the RW metric display resemblances between the meVSL and SMC models, distinctions arise in the Ricci curvature tensors due to the time-varying speed of light. These differences stem from the derivatives of the Christoffel symbols involving the changing speed of light concerning cosmic time $t$ (or the scale factor $a$). These derivatives depict distortions of shapes along geodesics in space. Once more, due to temporal fluctuations in the speed of light, correction terms impact both $R_{00}$ and $R_{ij}$
\begin{align}
R_{00} &= -\frac{3}{c^2} \left( \frac{\ddot{a}}{a} - \frac{\dot{a}^2}{a^2} \frac{d \ln c}{ d \ln a}  \right) \,, 
R_{ij} = \frac{\gamma_{ij}}{c^2} a^2 \left( 2 \frac{\dot{a}^2}{a^2} + \frac{\ddot{a}}{a} + 2 k \frac{c^2}{a^2} - \frac{\dot{a}^2}{a^2} \frac{d \ln c}{ d \ln a}  \right) \label{RijApp} \,. 
\end{align}
One can trace the Ricci tensors to determine the Ricci scalar 
\begin{align}
R &= \frac{6}{c^2} \left( \frac{\ddot{a}}{a} + \frac{\dot{a}^2}{a^2} + k \frac{c^2}{a^2} - \frac{\dot{a}^2}{a^2} \frac{d \ln c}{ d \ln a}  \right) \label{RmpApp} \,,
\end{align}  
where the VSL effect becomes apparent in the final term. By working through these equations, one can derive the components of the Einstein tensor
\begin{align}
G_{00} &= \frac{3}{c^2} \left[ \frac{\dot{a}^2}{a^2} + k \frac{c^2}{a^2} \right] \quad ,  \quad
G_{ij} = - \frac{\gamma_{ij} a^2}{c^2} \left[ \frac{\dot{a}^2}{a^2} + 2 \frac{\ddot{a}}{a} + k \frac{c^2}{a^2} - 2 \frac{\dot{a}^2}{a^2} \frac{d \ln c}{d \ln a} \right] \label{G00Gii} \,.
\end{align}

\subsection{Bianchi identity and cosmological evolution of rest mass}
\label{subsec:BI}

Even with the incorporation of time-varying terms for the speed of light in the Einstein tensors of VSL models, depicted in Eq. ~\eqref{G00Gii}, it can be shown that these models retain the characteristic that the covariant derivatives of both the Einstein tensors $G_{\mu\nu}$ and the metric $g_{\mu\nu}$ remain zero
\begin{align}
\nabla^{\nu} G_{\mu\nu} = 0 \quad , \quad \nabla^{\nu} g_{\mu\nu} = 0 \label{BI} \,.
\end{align}
This essential characteristic is known as the Bianchi Identity (BI).  By utilizing the BI along with the constancy of the Einstein gravitational constant $\kappa$, it is possible to derive the local conservation law for energy and momentum, as illustrated in Eq. \eqref{tEFEmpApp}
\begin{align}
\nabla^{\nu} \left( \kappa T_{\mu\nu} \right) = \kappa \nabla^{\nu} T_{\mu\nu} = 0 \quad \Rightarrow \quad \sum_{i} \left[ d \ln \left( \rho_i c^2 \right) + 3 \left( 1+ \omega_i \right) d \ln a \right] = 0  \label{CEM} \,,
\end{align}
where $\omega_i = P_{i}/(\rho_i c^2)$ denotes the equation of state (EOS) for the $i$-component. Consequently, one can infer the cosmological evolution of the mass density for the $i$-component in various models as 
\begin{align}
\rho_i = \rho_{i0} \left( c_0^2/c^2 \right)  a^{-3(1+\omega_i)} = \begin{cases}  \rho_{i0} a^{-3(1+\omega_i)} & \textrm{SMC} \\ 
\rho_{i0} \left( \frac{f(a)}{f(1)a} \right)^2 a^{-3(1+\omega_i)} & \textrm{VSL} \\ 
\rho_{i0} a^{-\frac{b}{2}} a^{-3(1+\omega_i)} & \textrm{meVSL} \end{cases} \label{rhoi} \,,
\end{align}
where we denote the subscript $0$ as the present values of corresponding quantities and utilize Eq.~\eqref{dl3D}.  Eq. ~\eqref{rhoi} is valid as long as there is no interaction between components. As usual, we consider no interaction between different components, such as radiation, matter, and dark energy. Additionally, we set $a_0 = 1$. By interpreting Eq.~\eqref{rhoi} as the dynamic cosmological rest mass within VSL models, we can achieve a coherent derivation of a fluid's covariant rest mass energy with a constant eos. Therefore, understanding the cosmological evolution of rest mass is crucial for establishing consistent VSL models
\begin{align}
E= m c^2 = m_0 c_0^2 \quad \textrm{for\,\,SMC\,,VSL\,,and\,meVSL} \,. \label{mc2}
\end{align}

\section{Friedmann equation and Huuble tension}
\label{sec:FriEq}

We express the $00$-component of the EFEs as 
\begin{align}
&\frac{\dot{a}^2}{a^2} + k \frac{c^2}{a^2} - \frac{\Lambda c^2}{3} = \frac{8 \pi G}{3} \sum_{i} \rho_i \quad \Rightarrow \quad 
\frac{\dot{a}^2}{a^2} \equiv H^2 = \begin{cases} H_{(\textrm{SMC})}^{2} & \textrm{SMC} \\ H_{(\textrm{SMC})}^{2} \left( \frac{a}{f(a)} \right)^2 & \textrm{VSL} \\ H_{(\textrm{SMC})}^{2}a^{\frac{b}{2}} & \textrm{meVSL} \end{cases} \,,  \nonumber \\
&\textrm{where}  \quad H_{(\textrm{SMC})}^{2} \equiv \left[ \frac{8 \pi G_0}{3} \sum_{i} \rho_{i0} a^{-3(1+\omega_i)} + \frac{ \Lambda \tilde{c}_0^2}{3} - k \frac{\tilde{c}_0^2}{a^2} \right] \equiv H_0^2 E_{(\textrm{SMC})}^{2} \label{G00} \,.
\end{align}
In the context of the VSL (meVSL) model, the expansion rate of the Universe denoted as $H$, incorporates an additional factor of $a/f(a)$ ($a^{b/4}$) compared to the expansion rate of the SMC,  $H^{(\textrm{SMC})}$. Consequently, the present-day values of the Hubble parameter coincide for both the SMC and VSL (meVSL) models. However, a notable difference arises in the Hubble parameter's value between VSL (meVSL) and SMC within an expanding Universe framework. This straightforward observation offers potential insights into addressing the Hubble tension \cite{Lee:2020zts}.

\subsection{Hubble radius}
\label{subsec:HR}

The primary aim of earlier VSL models has been to offer an alternative framework for elucidating cosmic inflation. These models introduce the concept of a diminishing comoving Hubble radius over time, indicated by the condition $d(c/aH)/dt < 0$. The evolution of the comoving Hubble radius in VSL models relies on equations \eqref{dl3D} and \eqref{G00}.  Upon closer examination of equation \eqref{HR},  it becomes apparent that the Hubble radius in the context of VSL models is equivalent to that of the SMC
\begin{align}
\frac{\tilde{c}}{a H} = \frac{\tilde{c}_0}{a H^{(\textrm{SMC})}} \label{HR} \,.
\end{align}
This raises a crucial consideration: to maintain the viability of a VSL model, it is imperative to avoid postulating variations solely at the speed of light. There is a concern that the Hubble radius could align with that of the SMC, potentially compromising the uniqueness of the proposal.

\subsection{Local physics laws}
\label{subsec:LPL}

In the meVSL model, we take into account local thermodynamics, energy conservation, and other local physical phenomena including electromagnetism. The covariance of Maxwell's equations into the meVSL model results in the cosmological time evolutions of permeability, permittivity, and electric charge. These factors also lead to the temporal evolution of various physical constants and quantities, as outlined in Table\ref{tab:table-2}. 

\begin{table}[htbp]
\caption{Summary for cosmological evolutions of physical constants and quantities of the meVSL model. These relations satisfy all known local physics laws, including special relativity, thermodynamics, and electromagnetic force \cite{Lee:2020zts,Lee:2023FoP,Lee:2022heb}.}
\label{tab:table-2}
\begin{adjustbox}{width=\columnwidth,center}
\begin{tabular}{|c||c|c|c|}
	\hline
	local physics laws & Special Relativity & Electromagnetism & Thermodynamics \\
	\hline \hline
	quantities & $\tilde{m} = \tilde{m}_0 a^{-b/2}$ & $\tilde{e} = \tilde{e}_0 a^{-b/4}\,, \tilde{\lambda} = \tilde{\lambda}_0 a \,, \tilde{\nu} = \tilde{nu}_0 a^{-1+b/4}$ & $\tilde{T} = \tilde{T}_0 a^{-1}$ \\
	\hline
	constants & $\tilde{c} = \tilde{c}_0 a^{b/4} \,, \tilde{G} = \tilde{G}_0 a^{b}$ & $\tilde{\epsilon} = \tilde{\epsilon}_0 a^{-b/4} \,, \tilde{\mu} = \tilde{\mu}_0 a^{-b/4} $ & $\tilde{k}_{\textrm{B} 0} \,, \tilde{\hbar} = \tilde{\hbar}_0 a^{-b/4}$ \\
	\hline
	energies & $\tilde{m} \tilde{c}^2 = \tilde{m}_0 \tilde{c}_0^2$ & $\tilde{h} \tilde{\nu} = \tilde{h}_0 \tilde{\nu}_0 a^{-1}$ & $\tilde{k}_{\textrm{B}} \tilde{T} = \tilde{k}_{\textrm{B}} \tilde{T}_0 a^{-1}$ \\
	\hline
\end{tabular}
\end{adjustbox}
\end{table}


\section{Observational constraints}
\label{sec:Obs}

Identifying observational methods that can be compared with the SMC is crucial because of the limited observational techniques available for investigating the meVSL model. In this section, we briefly describe such observational methods.

\subsection{Cosmic distance duality relation}
\label{subsec:CDDR}

Etherington's theorem states that the area distance of a galaxy and that of an observer are proportional, with the redshift factor $(1+z)$, under the assumption of geometric invariance when roles are interchanged between the source and the observer \cite{Etherington:1933pm}. This reciprocity theorem, derived from the geodesic deviation equation, holds in any spacetime where photons follow null geodesics and the geodesic deviation equation is valid. If photon conservation is assumed, the cosmic distance duality relation (CDDR) can be derived from this reciprocity theorem, linking these area distances to angular and luminosity distances \cite{Ellis:1998ct,Ellis:2007grg}. Consequently, the CDDR offers a means to test the validity of the SMC, regardless of its spacetime background.

The angular diameter distance and the luminosity distance are both essential in determining cosmological parameters in modern cosmology. The validity of the CDDR underpins this analysis, making it crucial to investigate its accuracy. Various mechanisms could lead to the violation of one or more conditions of the CDDR. The relation is expressed as
\begin{align}
\frac{d_{L}}{d_{A}} (1 + z)^{-2} = 1 \label{cCDDR} \,,
\end{align}
where $d_{L}$ represents the luminosity distance and $d_{A}$ denotes the angular diameter distance. Testing this relationship involves measuring sources with known intrinsic luminosities (standard candles) and intrinsic sizes (standard rulers). While ideally, these measurements should be model-independent, practical limitations necessitate reliance on cosmological observations based on specific models. Several tests of the CDDR have been conducted using various astrophysical and cosmological observations.

The validity of the CDDR has been scrutinized using angular diameter distances derived from baryon acoustic oscillations (BAO) in conjunction with luminosity distances obtained from Type Ia supernovae (SNIa) \cite{More:2008uq,Nair:2012dc,Wu:2015prd,Ma:2016bjt,Martinelli:2020hud}.  There have been observational constraints on VSL models from CDDR \cite{Holanda:2012ia,Qi:2014zja,Salzano:2014lra,Lee:2020zts,Lee:2021xwh,Rodrigues:2021wyk,Cuzinatto:2022mfe}. We conducted a maximum likelihood analysis for the meVSL model on combined datasets, revealing that certain results suggest a $1$-$\sigma$ deviation from the standard CDDR based on current data \cite{Lee:2021xwh}. However, upon employing different priors for certain cosmological parameters, the current dataset aligns with the SMC exhibiting no deviation from the expected CDDR. Therefore, it is necessary to obtain more accurate data to thoroughly investigate any potential deviations from the established CDDR. Additionally, our findings reaffirm the viability of the meVSL model, providing an additional constraint on the parameter $b$ from CDDR, supplementing previous constraints obtained from alternative investigations. 

\subsection{Cosmic chronometer}
\label{subsec:CC}

The Cosmic Chronometer (CC) method involves observing two passively evolving galaxies, typically elliptical galaxies, assumed to have formed at the same cosmic epoch but observed at different redshifts, as outlined in \cite{Jimenez:2001gg}. This approach offers a model-independent means of measuring the Hubble parameter, $H(z)$, as a function of redshift. The difference in their redshifts, $dz$, is derived from spectroscopic surveys with high precision ($\sigma_z \leq 0.001$). Subsequently, the expansion rate, or the Hubble parameter $H(z)$, is determined from the differential age evolution of the Universe $\Delta t$ within a given redshift interval ($dz$) as expressed 
\begin{align}
H(z) \equiv \frac{\dot{a}}{a} &= - \frac{1}{1+z} \frac{dz}{dt} \approx - \frac{1}{1+z} \frac{\Delta z}{\Delta t} = H(z)^{(\textrm{SM})} (1+z)^{-b/4} \nonumber \\ &= H_0 E(z)^{(\textrm{SMC})} (1+z)^{-b/4} \label{HzCC} \,,
\end{align}
where $E^{(\textrm{SM})}$ is defined in Eq.~\eqref{G00}, and the differential redshift-time relation ($dz/dt \approx \Delta z/\Delta t$) is assumed measurable. Various methods exist for measuring $\Delta t$, including predicting its age based on the chemical composition of a stellar population or utilizing spectroscopic observables like the $4000$ \AA break, known to be linearly related to the age of the stellar population \cite{Moresco:2010wh}. Unlike many cosmological measurements that rely on integrated distances, the CC method determines the expansion rate $H(z)$ as a function of the redshift-time derivative $dz/dt$, making it a potent tool for testing different cosmological models \cite{Wei:2016ygr,Ratsimbazafy:2017vga,Wei:2019uss,Moresco:2020fbm,Vagnozzi:2020dfn,Dhawan:2021mel,Borghi:2021zsr,Borghi:2021rft,Banerjee:2022ynv,Jalilvand:2022lfb,Asimakis:2022jel,Kumar:2022ypo,Li:2022cbk}. This method proves particularly valuable for investigating VSL models\cite{Rodrigues:2021wyk}.  We have performed both minimum $\chi^2$ analysis and maximum likelihood analysis using the most recent CC data to constrain the parameter $b$ of the meVSL model. Our findings indicate that the precision of the current CC data is insufficient to distinguish between meVSL and SMC \cite{Lee:2023rqv}.

\subsection{Time dilations in SNe Ia}
\label{subsec:TDSNeIa}

The luminosity curve (LC) of an SN serves as a comprehensive record of its brightness evolution over time, revealing crucial insights into its behavior. It begins with a phase known as the ``pre-maximum", characterized by a rise in brightness leading up to its peak luminosity, followed by a subsequent decline. This LC shape holds valuable information about the SN, particularly for SNe Ia which serve as standard candles in cosmology. Through LC analysis, astronomers can ascertain key parameters such as peak luminosity, time taken to reach maximum brightness, and the rate of decline. It aids in the classification of SNe and enhances our understanding of their energetics, composition, and explosion mechanisms.

Comparing LCs across varying distances enables the investigation of cosmic expansion and TD, contributing to significant discoveries like the accelerated expansion of the Universe and the presence of dark energy. Wilson's method involves comparing the LCs of nearby and distant SNe \cite{Wilson:39}. Distant SNe exhibit TD, causing their LCs to appear stretched compared to nearby counterparts due to the light travel time through space. By analyzing TD effects in LCs, researchers can deduce the time taken for light from SNe at different distances to reach the observer.

This information is pivotal for studying the Universe's expansion rate and testing various cosmological models. In practice, this entails collecting data on supernova brightness and evolution across different redshifts, fitting mathematical models to their LCs, and comparing observed TD with theoretical predictions. 

We have derived a formula describing the TD of distant objects within the framework of the meVSL model, given by $T(z) = T_0 (1 + z)^{1 - \frac{b}{4}}$. By analyzing TD data from 13 high-redshift SNe Ia \cite{Blondin:2008mz}, we have determined the best-fitting values for the exponent $b$ to be $b = 0.198 \pm 0.415$ at the $1$-$\sigma$ confidence level. However, this result is less precise compared to that obtained using CC \cite{Lee:2023rqv}.

Our analysis indicates that the current data are consistent with both the expectations of the standard cosmological model and those of the meVSL model. Thus, based on the time dilation data from SNe, we are unable to distinguish between the meVSL model and the SMC \cite{Lee:2023ucu}.

\subsection{Cosmpgraphy}
\label{subsec:Cosmogra}

The method of cosmography uses the kinematic description of the evolution of the Universe depending solely on the cosmological principle, with a specific emphasis on the dynamics of cosmological expansion. As a model-independent framework, cosmography provides a flexible platform for managing cosmological parameters. This approach allows for a more generalized analysis, free from the constraints of preconceived models. Focusing predominantly on the later stages of the Universe's evolution, cosmography utilizes Taylor expansions tailored to the observable domain where $z \ll 1$, enabling the imposition of constraints on the present-day Universe. We provide the adaptation of late-time cosmography to accommodate meVSL models \cite{SLee:24CG}.

\section{Discussion}
\label{sec:Conc}

The $\Lambda$CDM cosmological model relies on the Robertson-Walker metric. To comprehend this model, we must examine the role of the cosmological principle and Weyl’s postulate in constructing this metric. Fundamental observers in this context are in free fall, and their proper time coincides with coordinate time because their spatial position remains constant along their worldlines. The surface of simultaneity for these observers establishes a link between coordinate and cosmic time. When defining fundamental observers as comoving with matter, we can define cosmic time utilized in the global metric and energy stress tensor. In the traditional Robertson-Walker metric, the constancy of the speed of light arises from a specific assumption of cosmological time dilation, rather than being a direct consequence of the cosmological principle and Weyl’s postulate. Without explicit physics laws governing time dilation, the speed of light in this metric can vary with cosmic time, akin to other physical quantities like mass density, temperature, and fundamental constants such as the Planck constant. It suggests the possibility of a varying speed of light with cosmic time. Essentially, the assumption that the speed of light remains constant for cosmic time in the traditional Robertson-Walker metric was an additional postulate introduced through the assumed cosmological time dilation. This theoretical concept implies that, unless experimentally validated, the speed of light, like other physical quantities, undergoes cosmological redshift in an expanding universe. This model does not contradict the original principles of the Robertson-Walker metric. Therefore, we should check its validity through observations \cite{Lee:2023rqv,Lee:2023ucu,SLee:24CG}.


\vspace{6pt}

\section*{Acknowledgments}
SL is supported by Basic Science Research Program through the National Research Foundation of Korea (NRF) funded by the Ministry of Science, ICT, and Future Planning (Grant No.  NRF-2017K1A4A3015188,  NRF-2019R1A6A1A10073079,  NRF-2022R1H1A2011281).  



\end{document}